\documentclass[twocolumn,showpacs,preprintnumbers,amsmath,amssymb]{revtex4}

\usepackage{graphicx}
\usepackage{dcolumn}
\usepackage{bm}

\begin{document}

\preprint{APS/123-QED}

\title{Universalities in One-electron Properties of Limit Quasi-periodic Lattices }

\author{Rihei Endou$^1$}
\email{rihei@cmpt.phys.tohoku.ac.jp}
\author{Komajiro Niizeki$^1$}\author{Nobuhisa Fujita$^2$}
\affiliation{$^1$Department of Physics, Graduate School of Science, Tohoku University, Sendai 980-8578, Japan}
\affiliation{$^2$Structural Chemistry, Arrhenius Laboratory, Stockholm University, 10691 Stockholm, Sweden}

\date{\today}

\begin{abstract}
We investigate one-electron properties of one-dimensional self-similar 
structures called limit quasi-periodic lattices. 
The trace map of such a lattice is nonconservative in contrast 
to the quasi-periodic case, and we can determine the structure of its attractor. 
It allows us to obtain the three new features of the present system:
1) The multi-fractal characters of the energy spectra are {\it universal}.
2) The supports of the $f(\alpha)$-spectra extend over the whole unit interval, $[0, \,1]$.
3) There exist marginal critical states.

\end{abstract}

\pacs{61.44.Br, 64.60.Ak, 71.23.Ft}
\keywords{Suggested keywords}
\maketitle

Since the discovery of a quasicrystal, deterministic and aperiodic systems 
have been attracted much attention \cite{St99}.
They are classified into the third group in addition to periodic systems and random systems. 
Although they include a wide range of diverse structures, 
a self-similar structure such as the Fibonacci lattice or the Penrose tiling is especially important 
because quasicrystals belong to materials of this type. 
In particular, a one-dimensional (1D) self-similar structure has the symmetry described by a semi-group, 
and the electronic state on it exhibits rich properties \cite{KKT, KO, KKL, HK}. 
The presence of multifractal energy spectra and self-similar wave functions 
is peculiar to such systems. 
Here we will add {\it universality} to the subject; 
it applies to limit quasiperiodic lattices which form an important class of self-similar structures. 

A 1D self-similar structure is produced with a substitution rule (SR) 
starting from two elements, $A$ and $B$ \cite{KAI}. 
It is taken to be an aperiodic lattice composed of two types of sites, and called a self-similar lattice (SSL).
A general form of the SR is written as $A \rightarrow \sigma_A(A, \,B), B \rightarrow \sigma_B(A, \,B)$, 
where $\sigma_A(A, \,B)$ and $\sigma_B(A, \,B)$ are words of the two letters or, 
equivalently, finite sequences of the two symbols. 
The SR can be specified by a combination, $(\sigma_A(A, \,B), \sigma_B(A, \,B))$, of two different words, 
and an infinite number of different SRs are formed by changing the combination. 
A simplest SSL is the Fibonacci lattice, whose SR is $(B, \,AB)$, 
while another one with the SR, $(B, \,BAAB)$, produces the mixed mean lattice (MM lattice) \cite{KAII}. 
To every SR, $(\sigma_A, \,\sigma_B)$, 
we can associate a Frobenius matrix, $M :=  \left ({a \atop c}\; {b \atop d}\right)$ 
with $a$ and $c$ (resp. $b$ and $d$) being the numbers of $A$ and $B$ 
in $\sigma_A(A, \,B)$ (resp. $\sigma_B(A, \,B)$); 
we shall called it the substitution matrix of the SR or of the relevant SSL. 
An SR is not uniquely determined by its substitution matrix 
because two or more choices are usually possible for the orders of the letters, 
A and B, \,in $\sigma_A$ and $\sigma_B$. 
If there are several locally-isomorphic classes of SSLs with a common substitution matrix, 
we may call them isomers. 
For example, two SSLs with SRs, $(B, ABBA)$ and $(B, ABAB)$, are isomers of the MM lattice.  
When an SSL has a global center of the reflection symmetry, 
it can be called a symmetric SSL and the relevant SR a symmetric SR \cite{TNO}. 
We may call an SR, $(\sigma_A, \,\sigma_B)$, palindromic 
if both $\sigma_A$ and $\sigma_B$ are palindromes, 
while we may call it quasi-palindromic if it is written with a palindromic SR, 
$(\sigma'_A, \,\sigma'_B)$, as $(\sigma'_AB, \,\sigma'_BB)$. 
Palindromic SRs and quasi-palindromic ones produce SSLs. 
If a substitution matrix is specified, at least one of the relevant isomers is a symmetric SSL. 
All the SRs presented above produce symmetric SSLs. 

We shall turn to the structure factor $S(Q)$ an SSL \cite{LGJJ, K93}. 
The set of all the SSLs is divided into two classes $\Pi$ and $\Gamma$; 
$S(Q)$ for the case $\Pi$ consists only of Bragg peaks, 
while the one for the case $\Gamma$ is singular continuous.
The class $\Pi$ is divided, further, into the three subclasses, i) $\Pi_{\rm I}$: Quasiperiodic, 
ii) $\Pi_{\rm II}$: Limit Quasiperiodic, and iii) $\Pi_{\rm III}$: Limit Periodic. 
Representatives of the three are the Fibonacci lattice, the MM lattice, 
and the period-doubling lattice, respectively.
Let $M$ be the substitution matrix of an SSL, ${\cal L}$, 
and $\tau \;(> 1)$ be the the Frobenius eigenvalue of $M$.
Then, a necessary and sufficient condition for ${\cal L}$ to belong to 
$\Pi_{\rm I}$ or $\Pi_{\rm II}$ is that the second eigenvalue $\tau'$ satisfies $|\tau'| < 1$. 
If it is satisfied, $\tau$ is a quadratic irrational called a Pisot number. 
Note that $\tau'$ is the algebraic conjugate of $\tau$ and $\tau \tau' = s$ with $s :=  \det M$; 
$M$ is unimodular only when $s = \pm 1$. 
It is known that ${\cal L}$ belongs to $\Pi_{\rm I}$ if $|s| = 1$ but to $\Pi_{\rm II}$ otherwise. 
For example, $\tau = (1+\sqrt{5})/2$ (resp. $\tau = 1+\sqrt{3}$) for 
the Fibonacci lattice (resp. the MM lattice). 
The set of all the wave numbers at the peaks of $S(Q)$ form an additive group, ${\cal M}$, 
which is called the Fourier module. 
The two subclasses $\Pi_{\rm I}$ and $\Pi_{\rm II}$ are distinguished by whether 
the number of the generators of ${\cal M}$ is two or infinite. 
Let $M$ be the substitution matrix of an SSL and 
let $\omega :=  (\tau - a)/b = c/(\tau - d)$ with $\tau$ being the Frobenius eigenvalue of $M$. 
Then, the Frobenius eigenvector of $M$ is given by the column vector ${}^{\rm t}(1, \,\omega)$. 
The Fourier module ${\cal M}$ of an SSL in $\Pi_{\rm II}$ is written as 
${\cal M} = (1 + \omega)^{-1}{\bf Z}\{\omega\}$ with 
${\bf Z}\{\omega\} :=  {\bf Z}[\omega] \cup 
\tau^{-1}{\bf Z}[\omega] \cup \tau^{-2}{\bf Z}[\omega] \cup \cdots$, 
where ${\bf Z}[\omega] :=  \{x+y\omega \,|\,x,y \in {\bf Z}\}$. 
The Fourier module is a dense set on the real axis.
Note that ${\cal M}$ is common among different isomers 
because ${\cal M}$ is uniquely determined by $M$. 
Incidentally, ${\bf Z}\{\omega\} = {\bf Z}[\omega]$ and 
${\cal M} = (1 + \omega)^{-1}{\bf Z}[\omega]$ for an SSL in ${\rm \Pi_{I}}$ 
becasue $\tau{\bf Z}[\omega] = {\bf Z}[\omega]$ for this case. 

Every set, say $\Pi$, of SSLs is divided into the symmetric 
subset $\Pi^{\rm s}$ and its complement $\Pi^{\rm a}$, 
where $\Pi^{\rm s}$ is composed only of symmetric SSLs. 
Previous researches on the one-electron states of SSLs 
are focused exclusively on the case of class $\Pi^{\rm s}$. 
More precisely, the subclass $\Pi_{\rm I}^{\rm s}$ is investigated extensively \cite{HK} but
$\Pi_{\rm II}^{\rm s}$ is investigated by nobody, 
while there are a considerable number of papers on the subclass $\Pi_{\rm III}^{\rm s}$ 
but they are still on a rudimentary stage (see, for example, \cite{KAII}). 
Therefore, the authors investigated in detail the case of the subclass $\Pi_{\rm II}^{\rm s}$. 
Though $\Pi_{\rm II}^{\rm s}$ appears a simple extension of $\Pi_{\rm I}^{\rm s}$, 
it turns out as shown below that its one-electron states exhibit 
a remarkably different character from the case of the latter. 
We shall call hereafter an SSL in $\Pi_{\rm II}^{\rm s}$ a limit quasi-periodic lattice (LQL). 

Self-similarity of an SSL allows us to use 
the trace map (TM) for researches of the electronic states on it \cite{KKT, KO}. 
The character of the TM as a non-linear map depends crucially on 
whether it is conservative or not \cite{KAII, RB, WWL}. 
It is known that SSLs with conservative TMs form a special subset, 
$\Pi_{\rm I}^{\rm c}$, of $\Pi_{\rm I}^{\rm s}$ \cite{TNO, WWL}. 
Hereafter, an SSL in $\Pi_{\rm I}^{\rm c}$ is called simply a conservative SSL (CSSL). 
Remember that an LQL is never conservative. 

\begin{figure}
\includegraphics[width=.90\linewidth]{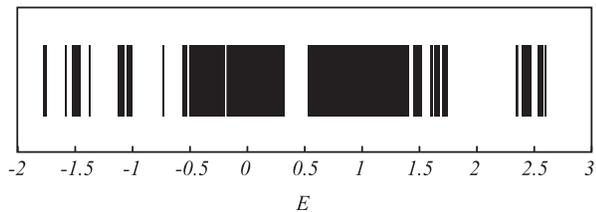}
\caption{The energy spectrum of the MM lattice with $\Delta :=  V_B - V_A = 1$} 
\label{fig.1}\end{figure}

Generic characters of one-electron states of an SSL can be 
investigated using the tight-binding Hamiltonian ${\cal H}$, 
each site-energy of which takes $V_A$ or $V_B$ depending on the type of the site \cite{HK}. 
The unit of energy can be so chosen that the transfer integral assumes $-1$. 
It was shown generally that the energy spectrum $\sigma = \sigma ({\cal H})$ 
is singular continuous \cite{HKS}. 
More precisely, it is a bounded closed set having an infinite number of gaps, 
and, moreover, the length of its every connected component vanishes; we can see it in Fig.1. 
This means that the integrated state density $H(E)$ per one site is a devil's staircase. 
Every gap of $\sigma$ can be specified by the height of the corresponding step of the staircase. 
Let $G_\sigma$ be the set formed by adding 0 and 1 to the set of the heights of all the steps. 
Then, it is a discrete subset of the unit interval $[0, \,1]$, and 
is determined by the gap labeling theorem \cite{BBG}.
Since $\sigma$ is a multifractal, its singularity, $\alpha = \alpha(E)$ with $E \in \sigma$, 
is a function of $E$. 
The distribution of $\alpha$ in $\sigma$ is characterized by the $f(\alpha)$-spectrum \cite{HK}. 
The support of $f(\alpha)$ is a closed interval, $[\alpha_{\rm min},\alpha_{\rm max}] \subset [0, \,1]$. 
For every $E$ in $\sigma$, we can define a function by 
$h (\varepsilon) :=  H(E + \varepsilon/2) - H(E - \varepsilon/2)$. 
If it satisfies the equation, $h (\varepsilon) \approx \mu h (\varepsilon/\lambda)$, 
for two parameters $\lambda = \lambda (E)$ and $\mu = \mu (E)$ chosen suitably, 
$E$ is a center of local self-similarity of $\sigma$. 
The singularity $\alpha$ of the center is given by $\alpha=\ln{\mu}/\ln{\lambda}$ \cite{KO, HK}. 
A remarkable feature of an SSL is that the set, $\sigma_{\rm ls}$, of 
all such centers is a countably infinite set \cite{KKL}. 
The set, $\Sigma :=  \{H(E) \,|\,E \in \sigma_{\rm ls}\}$ is an important object 
which characterizes $\sigma_{\rm ls}$. 

In the trace map formalism, a homomorphism is defined 
from ${\cal F}(A, \,B)$, the free semi-group generated by the two generators $A$ and $B$, 
into $SL(2, \,{\bf R})$, a group formed of $2\times 2$ unimodular matrices \cite{KAII, RB, AP}. 
If ${\hat A}$ and ${\hat B}$ are the images of $A$ and $B$ in the homomorphism, 
the image, $W({\hat A}, \,{\hat B})$, of a sequence $W(A, \,B)$ is called the transfer matrix 
associated with the sequence. 
We can form an infinite series of pairs, $\{A_n, \,B_n\}$, of sequences 
by using the recursion relation 
$\{A_{n+1}, \,B_{n+1}\} = \{\sigma_A(A_n, \,B_n), \sigma_B(A_n, \,B_n)\}$ 
supplemented by the initial condition, $\{A_1, \,B_1\} := \{A, \,B\}$. 
The corresponding pair, $\{{\hat A}_n, \,{\hat B}_n\}$, of transfer matrices 
satisfy a homologous  recursion relation. 
Let $\{{\hat A}, \,{\hat B}\}$ be such a pair at any generation 
and $W(A, \,B)$ any sequence. Then, ${\rm Tr}\,W({\hat A}, \,{\hat B})$ is a polynomial
with integral coefficients of the three variables, 
$x :=  {\rm Tr}\,{\hat A}$, $y :=  {\rm Tr}\,{\hat B}$, and $z :=  {\rm Tr}({\hat A}{\hat B})$. 
Thus, a map is defined from ${\cal F}(A, \,B)$ into ${\bf Z}[x, \,y, \,z]$.
Let $T_x$, $T_y$, and $T_z$ be the three polynomials defined as the images of the three sequences, 
$\sigma_A(A, \,B)$, $\sigma_B(A, \,B)$, and $\sigma_A(A, \,B)\sigma_B(A, \,B)$, respectively. 
Then, the trace map $T$ which yields a 3D dynamical system over a 3D phase space ${\bf R}^3$ 
is defined by ${\bf r} \rightarrow T {\bf r} :=  (T_x ({\bf r}), \,T_y ({\bf r}), \,T_z ({\bf r}))$ 
with ${\bf r} = (x, \,y, \,z)$ \cite{KKT, KAI, RB}. 
For example, $T{\bf r}$ assumes $(y, \,x, \,yz - x)$ for the Fibonacci lattice but 
\begin{eqnarray}
(y, \,xyz - x^2 - y^2 + 2, \,x(y^2 - 1)z - x^2y - (y^2 - 3)y) \label{TMMM}
\end{eqnarray}
for the MM lattice. 
Noncommutativity between the transfer matrices in a pair 
$\{{\hat A}, \,{\hat B}\}$ can be quantified by the polynomial defined by \cite{KKT}
\begin{eqnarray}
I :=  \frac{1}{2}{\rm Tr}[{\hat A}, \,{\hat B}]^2 = x^2 + y^2 + z^2 - xyz - 4, \label{Invariant}
\end{eqnarray}
which vanishes if they are commutative. 
It can be shown readily that $I({\bf r}_1) = (\Delta/2)^2$. 
If the subscript $n$ being suppressed at the moment is recovered, 
we may write ${\bf r}_{n+1} = T{\bf r}_n$. 
An orbit $O(E) :=  \{{\bf r}_n\,|\,n \in {\bf N}\}$ in the phase space 
is generated from the initial point ${\bf r}_1 = (V_A - E, \,V_B - E, \,(V_A - E)(V_B - E) - 2)$. 
An important property of the TM is that $T{\bf r}$ itself 
does not depend on the parameters $V_A$, $V_B$, and $E$. 

It can be shown generally that there exists a polynomial $P({\bf r})$ such that 
$[{\hat A}_{n+1}, \,{\hat B}_{n+1}] = P({\bf r}_n)[{\hat A}_n, \,{\hat B}_n]$ for a palindromic SR 
but $[{\hat A}_{n+1}, \,{\hat B}_{n+1}] = P({\bf r}_n)[{\hat A}_n, \,{\hat B}_n]{\hat B}_n$ 
for a quasi-palindromic one \cite{NE}. 
Moreover, $|P({\bf r})| \equiv 1$ for the conservative case but ${\rm deg}(P) \ge 1$ otherwise. 
We obtain $P = z$ for the MM lattice, for example. 
It yields the identity $I(T{\bf r}) \equiv [P({\bf r})]^2I({\bf r})$. 
Therefore, $I({\bf r})$ is an invariant of the TM only for the case of a CSSL 
but $I({\bf r})$ is a semi-invariant for the general case because 
the TM never changes its sign  \cite{KAI}. 

Let us assume that ${\hat A}_n$ and ${\hat B}_n$ for ${}^\exists n \in {\bf
N}$ are not commutative and, besides, that $P(x_n, y_n, z_n) = 0$ for some
$E \in \sigma$. 
Then,  ${\hat A}_m$, ${\hat B}_m$ are commutative for ${}^\forall m \ge n +
1$, and we can conclude that the relevant eigenstate will be extended as in
the case of a periodic system \cite{NY}.
Different extended states may be obtained from the roots of a similar
equation at a different generation.
Therefore, an SSL may have an infinite number of extended states unless it
is not an CSSL.  

The dynamical system defined by the TM, $T$, is characterized by its limit cycles (LCs). 
If a point in ${\bf R}^3$ belongs to an LC with period $p$, it is a fixed point of $T^p$ and 
an orbit starting from it is a pure cycle. 
A necessary and sufficient (N\&S) condition for $E\in \sigma$ to belong to 
$\sigma_{\rm ls}$ is that $O(E)$ falls on an LC. 
If this is satisfied, we can determine the scaling parameter $\lambda$ 
by a linear analysis of $T^p$ around the relevant fixed point, 
where $p$ is the period of the LC, 
so that the singularity $\alpha = \alpha (E)$ can be evaluated \cite{KO}. 
The wave function of the corresponding eigenstate is known to be asymptotically self-similar; 
the spatial ratio of the self-similarity is equal to $\tau^p$ \cite{HK}. 

Though $I(x, \,y, \,z)$ is not an invariant of the TM, $T$, of an LQL, 
the curved surface defined by the equation $I(x, \,y, \,z) = 0$ is an invariant surface. 
Its restriction, ${\bf S}$, into the cube $[-2, \,2]^3 \in {\bf R}^3$ is a closed surface, 
which is a ball with the tetrahedral point symmetry but has four cusps at 
$(2, \,2, \,2)$, $(2, \,-2, \,-2)$, $(-2, \,2, \,-2)$, and $(-2, \,-2, \,2)$ (see Fig.1c in \cite{RB}). 
A coordinate system can be introduced into ${\bf S}$ by 
$x = 2\cos {(2\pi u)}$, $y = 2\cos {(2\pi v)}$, and $z = 2\cos {[2\pi(u + v)]}$ with 
with $(u, \,v)$ being a variable on the 2D torus ${\bf T}^2 :=  {\bf R}^2/{\bf Z}^2$. 
Actually, ${\bf S}$ is doubly covered by ${\bf T}^2$ because 
$\pm (u, \,v)$ are mapped onto a single point on ${\bf S}$. 
The origin $(0, \,0)$ of ${\bf T}^2$ corresponds to $(2, \,2, \,2)$ in ${\bf R}^3$. 
The TM induces a 2D dynamical system (DS) over ${\bf T}^2$: $T(u, \,v) = (u, \,v)M$ 
with $M$ being the relevant substitution matrix. 
With these machineries together with a theory of algebraic number theory, 
we can determine all the cycles of the 2D DS, which will be presented elsewhere. 

For a periodic system ($\Delta = 0$), 
every 3D orbit of $T$ is confined to ${\bf S}$ from the very beginning. 
The initial state ${\bf r}_1$ of the orbit $O(E)$ with $E = -2\cos {2\pi \kappa}$ 
corresponds to $(u_1, \,v_1) = \pm(\kappa, \,\kappa)$ on ${\bf S}$, 
where $\kappa$ is the rationalized wave number of a plane wave state. 
It follows that $(u_n, \,v_n) = \pm(L_n^{(A)}\kappa, \,L_n^{(B)}\kappa)$ 
with $L_n^{(A)}$ and $L_n^{(B)}$ being the numbers of the letters in $A_n$ and $B_n$, respectively. 
A N\&S condition for the 2D orbit $\{(u_n, \,v_n) \,|\, n \in {\bf N}\}$ 
to converge on $(0, \,0) \in {\bf S}$ is equivalent to $\kappa \in {\cal M}$ \cite{LGJJ, K93}. 
On the other hand, the 2D orbit falls on a limit cycle with period $p$ 
if and only if $\{(u_{n+p} \mp u_n, \,v_{n+p} \mp v_n) \,|\, n \in {\bf N}\}$ converges on $(0, \,0) \in {\bf S}$. 
Using the fact that the ratios $L_{n+p}^{(A)}/L_n^{(A)}$ and $L_{n+p}^{(B)}/L_n^{(B)}$ 
tend to $\tau^p$ as $n$ goes to the infinity, we can show that 
a N\&S condition for that condition to be satisfied is that 
$(\tau^p - 1)\kappa \in {\cal M}$ or $(\tau^p + 1)\kappa \in {\cal M}$. 
It follows that a N\&S condition for the 2D orbit to fall on a limit cycle
is given by $\kappa \in {\bf Q}[\omega] - {\cal M}$ 
with ${\bf Q}[\omega] :=  \{u + v\omega \,|\, u, \,v \in {\bf Q}\}$. 
Thus, the behavior of the TM of a periodic system has been completely revealed. 
Since the energy spectrum of the periodic system is absolutely continuous, 
$\alpha(E)$ takes 1 \cite{alpha}. 
Two states with wave numbers $\pm \kappa$ degenerate in energy 
and the corresponding two orbits of the TM are identical, so that 
we can assume $0 \le \kappa \le 1/2$. 
The number $\kappa$ is nothing but the normalized value per one site of the number 
of the nodes in the relevant sinusoidal wave. 
Since $H(E)$ is related to $\kappa = \kappa (E)$ by the equation $H = 2\kappa (E)$, 
the asymptotic behavior of the TM at the energy $E$ is determined by the value of $H(E)$, 
which belongs to the interval $[0, \,1]$.

We will proceed to the case of a non-periodic system ($\Delta \ne 0$). 
Since an LQL has the reflection symmetry, every energy level has its own parity 
with respect to the center of symmetry. 
If $\Delta$ is changed adiabatically from zero to the present value, 
every energy level will change continuously 
because the degeneracies being present at $\Delta = 0$ does not matter if the parity is specified. 
It follows that $\kappa$ remains to be the normalized number of the nodes. 
Then, the asymptotic behavior of the TM will be determined by the value $H = H(E)$, 
and the rule is the same as what is described above for the periodic system. 
Thus, we can conclude that $\Sigma = {\bf Q}[\omega] \cap [0, \,1]$, 
which is a similar result to the one obtained in \cite{KKL} for the case of the Fibonacci lattice.
 A remarkable property of the TM of an LQL is its {\it universality}. 
It is based on the two points:
1) The TM does not explicitly depend on the parameters $\Delta$ and $E$. 
2) It is essentially of 3D character, and not confined to a 2D manifold as in the case of a CSSL. 
Thus, the attractor is independent of the value of $\Delta$, and 
identical to that of the periodic system. 
It follows that the $f(\alpha)$-spectrum of $\sigma$ is universal. 
Moreover, we can conclude that $\alpha = 1$ 
for every center of local self-similarity of $\sigma$ \cite{alpha}. 
For example, a 3D orbit for the MM lattice converges to 
a 1-cycle if $\kappa = 1/3$ , and the scaling parameter $\lambda$ 
determined by a linear analysis of $T$ around the relevant fixed point, 
$(-1, \,-1, \,-1)$, is found exactly equal to $\tau$, so that $\alpha = 1$. 
As a consequence, we obtain $\alpha_{\rm max} = 1$. 

From a brief perturbative consideration, we may expect that 
$G_\sigma = {\cal M} \cap [0, \,1]$, which is guaranteed by the gap-labeling theorem \cite{BBG}. 
The gap labeled by $H \in G_\sigma$ is bounded by two energy levels located at band edges. 
However, the corresponding fixed point $(2, \,2, \,2)$ of the TM is a singularity of ${\bf S}$, 
and we cannot conclude that $\alpha = 1/2$ for these two energy levels. 
Surprisingly, these states are actually marginal critical states, 
which we can prove by a similar technique to the one used in \cite{FN}. 
It follows that $\alpha_{\rm min} = 0$.
This and the result $\alpha_{\rm max} = 1$ 
are consistent with that the $f(\alpha)$-spectrum is universal 
because there is no reason why the value of $\alpha_{\rm min}$ or $\alpha_{\rm max}$ 
must take an odd value. 
The broadness of the support of the $f(\alpha)$-spectrum means 
that the energy spectrum of an LQL is more inhomogeneous than that of a CSSL, 
which one can feel if he compares Fig.1 with the energy spectrum of, say, the Fibonacci lattice \cite{HK}. 
We show in Fig.2 the $f(\alpha)$-spectrum of the energy spectrum in Fig.1. 

\begin{figure}
\includegraphics[width=.80\linewidth]{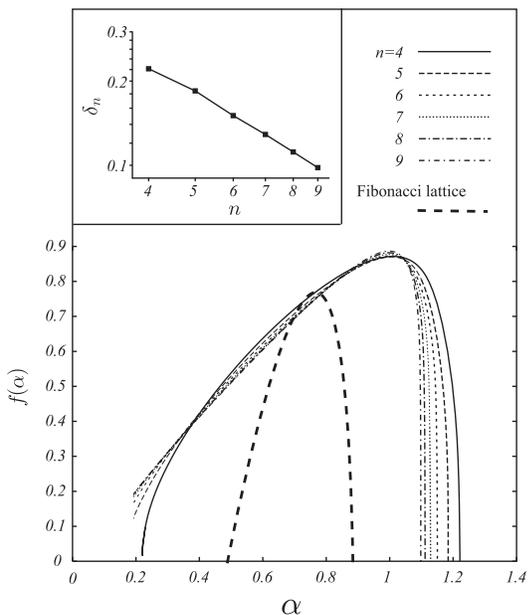} 
\caption{The $f(\alpha)$-spectra of the energy spectra 
of periodic approximants for the MM lattice. 
The data from the 4-th to the 9-th generations are used. 
The data in the region $\alpha < .2$ are cut because their accuracies are insufficient 
on account of the presence of marginal critical states. 
The inset is a log-log plot of $\delta_n := \alpha_{\rm max}^{(n)} - 1$ versus the generation number $n$. 
The asymptotic linearity of the plot concludes that $\delta_n$ 
tends to zero in an inverse power law for $n$ as $n$ goes to the infinity. 
The $f(\alpha)$-spectra of the Fibonacci lattice is superposed for a comparison. 
} 
\label{fig.2}\end{figure}

If there are several isomers, their TMs are different. 
Their $f(\alpha)$-spectra belong, however, to a common universality class 
because each universality class of the $f(\alpha)$-spectrum 
of the LQL can be specified by the substitution matrix. 

The TM derived from a CSSL is a 2D map on the curved surface $I(x, \,y, \,z) = (\Delta/2)^2$. 
Since this surface depends on the value of $\Delta$, the $f(\alpha)$-spectrum as well as 
$\alpha_{\rm min}$ and $\alpha_{\rm max}$ are not universal. 
This is consistent with the known fact that $\alpha_{\rm min} > 0$ and $\alpha_{\rm max} < 1$ 
for every CSSL (see Fig.2) \cite{KKT}. 
The SSLs in the set $\Pi_{\rm I}^{\rm c}$ are actually minorities in $\Pi_{\rm I}^{\rm s}$, 
and one-electron properties of SSLs in $\Pi_{\rm I}^{\rm s} - \Pi_{\rm I}^{\rm c}$ 
has never been investigeted yet. 
The present theory is basically applicable to this case; 
the only modification needed is to replace the Fourier module by
${\cal M} = (1 + \omega)^{-1}{\bf Z}[\omega]$. 
Thus, among isomers belonging to $\Pi_{\rm I}^{\rm s}$, 
only the CSSL member is nonuniversal. 
Incidentally, the SR, $(ABA, \,ABABA)$, produces a CSSL but its isomer, $(ABA, \,BAAAB)$, does not. 

Although no substance which takes a structure represented by an LQL 
as its thermodynamically stable phase is known, 
such a structure is realizable as an artificial super-lattice. 
Since a transfer matrix can be used also to a 1D Schr\"{o}dinger equation, 
the results of this letter apply to the electronic state in such an artificial super-lattice as well\cite{AC}. 
Furthermore, it is basically applicable to a propagation in a stratified substance 
of an ultrasonic wave, an electromagnetic wave (light), and a spin wave as well. 
If, for example, a lower marginal critical state is used as a channel of light, 
the speed of light will fall dramatically. 

In conclusion, one-electron properties of limit quasi-periodic lattices 
exhibits a universality in contrast to the case of quasi-periodic ones, 
and a universality class is specified solely by the substitution matrix.

One of the authors (K. N.) is grateful to Y. Morita.


\begin{thebibliography}{99}
\bibitem{alpha} However, $\alpha = 1/2$ for the case where $E$ is located at a band edge.
\end{thebibliography}
\end{document}